\documentclass[twocolumn,prl]{revtex4}

\usepackage[pdftex]{graphicx}
 \usepackage{pdfpages}
\usepackage{dcolumn}
\usepackage{bm}
\usepackage{multirow}
\usepackage{color}
\usepackage{xcolor}
\usepackage[normalem]{ulem}
\usepackage{amsmath}
\usepackage{gensymb}
\usepackage[colorlinks=true]{hyperref}
\newcommand{\blue}[1]{\textcolor{black}{#1}}

\begin{document}

\title{Thermal assisted transport of biexcitons in monolayer WSe$_2$}

\author{Dorian Béret$^{1}$}
\thanks{These two authors contributed equally}
\author{Louka Hemmen$^{1}$}
\thanks{These two authors contributed equally}
\author{Vishwas Jindal$^{1}$}
\author{Sreyan Raha$^{1}$}
\author{Thierry Amand$^1$}
\author{Delphine Lagarde$^1$}
\author{Andrea Balocchi$^1$}
\author{Cédric Robert $^1$}
\author{Hélène Carrere$^1$}
\author{Xavier Marie$^1$}
\author{Pierre Renucci$^1$}\
\author{Laurent Lombez$^1$}
\email{laurent.lombez@insa-toulouse.fr}

\affiliation{$^1$Universit\'e de Toulouse, INSA-CNRS-UPS, LPCNO, 135 Avenue Rangueil, 31077 Toulouse, France}

\begin{abstract} 
Studies of excitonic transport in transition metal dichalcogenide monolayers have attracted increasing interest in recent years in order to develop nano-optoelectronic devices made with 2D materials. These studies began with low to moderate optical excitation regimes, and more recently have focused on high injection regimes where nonlinear effects appear. This article is focused on the transport of biexcitons by spatially and temporally resolved photoluminescence spectroscopy at high excitation flux. The study is carried out on a high-quality $WSe_2$ monolayer encapsulated in hexagonal boron nitride. The results show that a Seebeck current affects transport in connection with the presence of hot biexcitons. In particular, we observe the formation of spatial rings, also called halos, which have been observed in other excitonic gases. These results tend to generalize the importance of high-energy populations in excitonic transport in TMD, even for complex and heavy excitonic particles.
\end{abstract}

\maketitle

\section{Introduction} 
Excitonic transport in transition metal dichalcogenide (TMD) monolayers has attracted considerable interest over the last decade\cite{perea-causin_exciton_2022,lee_recent_2023}. However, there are still many new physical aspects to study due to the wealth of excitonic species available. Indeed, while neutral excitons (intra- or interlayer) dominate the optical response in TMD monolayers and heterostructures at low or moderate carrier densities, it is necessary to consider higher-order multi-particle complexes that appear when increasing the doping concentration or the optical excitation power. \\
Moreover, the transition to higher exciton density regime could lead to unusual transport mechanism. For example, the increase in excitation density could lead to a hot excitonic population generating a Seebeck current which is a thermal current-assisted transport regime. This phenomenon has been quantified in classical 2D structures such as quantum wells \cite{gibelli_optical_2016,vezin_optical_2024}. In TMD, this phenomenon is highlighted by the appearance of a spatial halo profile \cite{perea-causin_exciton_2019, zipfel_exciton_2020,uddin_enhanced_2022,park_imaging_2021}.
Furthermore, by increasing the excitation density, an extremely efficient excitonic transport regime could emerge, the so-called hydrodynamic regime, where excitons behave like a superfluid \cite{fogler_high-temperature_2014,yu_giant_2020,aguila_ultrafast_2023}. For example, exciton scattering over several tens of micrometers has been reported in a $MoS_2$ monolayer at high excitation power\cite{aguila_ultrafast_2023}.\\
Here we focus on a $WSe_2$ monolayer encapsulated in hexagonal boron nitride ($hBN$), where we study the transport of multi-particle complexes at high excitation power. At high excitation density, we observe a non-linear transport regime assisted by a thermal gradient originating from hot biexcitons gases. We quantify the thermalization dynamics and decorrelate lattice temperature heating from the excitonic gas heating. This study focused on multi-exciton complexes confirms the key role played by hot populations in transport in TMD.\\

\section{Experimental results}

\subsection*{Spectral analysis} 
We first measured low-temperature photoluminescence spectra on our WSe$_2$ sample. 
\blue{We have fabricated van der Waals hBN/WSe$_2$/hbN heterostructures made of an exfoliated ML-WSe$_2$ embedded in high quality hBN crystals using a dry stamping technique\cite{cadiz_excitonic_2017}. The layers are deterministically transferred on top of a SiO$_2$/Si using PDMS substrate. Each layer deposition is followed by vacuum annealing at $340^\circ C$ for 6$~h$ under a pressure of $10^{-6}~mbar$ to ensure spatial homogeneity.}

\begin{figure*}
    \includegraphics[width=1\linewidth]{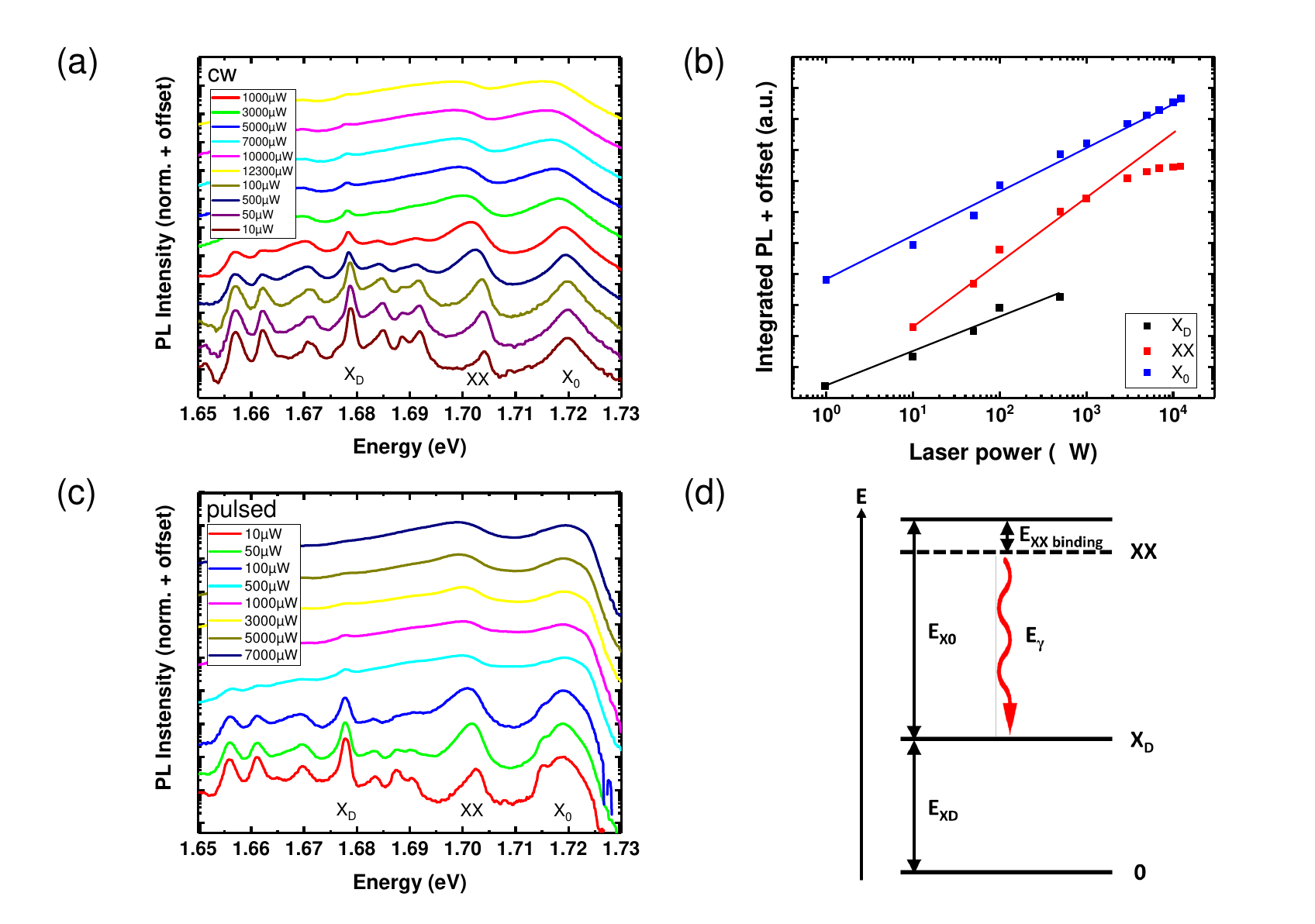}
    \caption{(a) PL spectra of a $WSe_2$ monolayer as a function of excitation power (cw-633nm laser). Spectra are \blue{normalized to the neutral exciton maximum intensity and shifted vertically to ease the reading}. (b) Double logarithmic plot of integrated PL intensity versus cw excitation power for $X_0$ (blue circles), $XX$ (red circles) and $X_D$ (black circles). The lines represent $I^{PL}\propto P^\alpha$ with $\alpha = 2.1$ for $XX$, $\alpha = 1.4$ for \blue{$X_0$} and $\alpha = 1.15$ for \blue{$X_D$}. Note that beyond $500~\mu W$ it is no longer possible to isolate the dark exciton into the PL spectrum. (c) PL spectra  as a function of excitation average power (pulsed-700nm laser). Spectra are shifted vertically for ease of reading. (d) Scheme of the excitonic energy levels involved in the radiative recombination process.}
    \label{fig:1}
\end{figure*}

Figure \ref{fig:1} (a) presents the evolution of PL spectra under a progressive increase of HeNe excitation laser ($1.96~eV$) power at $4~K$. \blue{In all experiments we use a microscope objective with a numerical aperture of 0.6.} We observed the presence of the bright neutral exciton at $1.720\;eV$ and the dark exciton at $1.679\;eV$ in agreement with the literature \cite{zhang_magnetic_2017,zhou_probing_2017,ye_efficient_2018}. By increasing the excitation power, a peak emerges at about $20\;meV$ below the bright exciton peak. This emission is attributed to the biexciton ($XX$)\citep{you_observation_2015,barbone_charge-tuneable_2018,li_revealing_2018,ye_efficient_2018,steinhoff_biexciton_2018}. According to literature, the biexciton is likely to be composed of a bright exciton and a dark exciton \blue{since this configuration minimizes exchange interactions between identical carriers}\cite{ye_efficient_2018}. The dark component of the biexciton is \blue{also long-lived, enabling the biexciton to be formed efficiently, as this population does not limit the biexciton formation.} At higher powers, the PL spectrum is largely dominated by neutral exciton and biexciton emissions.

Figure \ref{fig:1} (b) shows the integrated PL $I^{PL}$, defined as the area below the PL peak, as a function of laser excitation power $P$. By fitting the power dependence by a $I^{PL}\propto P^\alpha$ law, we deduce, via the $\alpha$ coefficient, a super-linear behavior of the bright exciton ($\alpha= 1.4$) and dark exciton ($\alpha= 1.17$) as well as a super-quadratic behavior of the biexciton peak ($\alpha= 2.3$). This is measured in the power range between $1~\mu W$ and $1~mW$ on a laser spot of about $1~\mu m^2$, beyond which the rise of the integrated PL area begins to slow down. These results are consistent with the literature \cite{you_observation_2015,barbone_charge-tuneable_2018,li_revealing_2018,ye_efficient_2018}, the near-quadratic power dependence of the biexciton suggests that it is formed from two excitons, each possessing linear (or close to linear) behavior with excitation power \cite{barbone_charge-tuneable_2018,li_revealing_2018,ye_efficient_2018}. By \blue{considering} dark and bright excitonic species as well as the biexciton formation from these two species , it has been shown that, for high biexciton generation densities, the power dependence becomes linear for excitation powers above $1~mW/\mu m^2$ in agreement with the observed slow down of \blue{the rise of }the PL intensity with $P$ \cite{ye_efficient_2018}.

Moreover, by looking at the evolution of the PL peak position as a function of power, we see a spectral shift towards lower energy (i.e. red shift) for the neutral exciton and biexciton lines (see Supplementary Material SM, Fig S1).  This energy shift could be attributed to an increase of the lattice temperature. According to the literature, a $5~meV$ shift would \blue{approximately} correspond to a $50~K$ temperature rise of the lattice  \cite{park_imaging_2021}.
The evolution of PL spectra as a function of power also shows significant line broadening.  This broadening is asymmetrical and pronounced at low energies. If we reach higher excitons density by using a $1.5~ps$ Ti:Sa laser pulse at $700~nm$ excitation wavelength, the spectral changes are even more pronounced on the two dominant peaks \blue{corresponding to} the exciton and biexciton (see Fig. \ref{fig:1} (c)). Studies carried out on the negative trion in $WSe_2$ show similar behavior. Concerning the trion emission, this broadening is explained by the existence of a high-energy trionic population that loses its kinetic energy through energy exchange with an electron gas, the so-called recoil effect involving two particles of different masses \cite{esser_photoluminescence_2000,zipfel_electron_2022,bauer_optical_2013}. This process was also mentioned to analyze the heating of a trion population in $MoSe_2$ \cite{park_imaging_2021}.
In our case, this spectral broadening is clearly seen in PL spectra at high flux. \blue{In this configuration we estimate a photogenerated carrier density varying from $10^{10}~cm^{-2}$ per pulse at low power to  $10^{14}~cm^{-2}$ for $10~mW$ which, in high exciton regime, might reach the Mott transition (around 2 to 3 $10^{13}~cm^{-2}$) in the first picoseconds  \cite{klingshirn_semiconductor_2012,lin_many-body_2019,steinhoff_exciton_2017,kudlis_modeling_2021,radisavljevic_mobility_2013,chernikov_electrical_2015}. Further discussion can be found in the Supplementary Material, section VIII.} These excitation conditions will be used in the following to carry out time-resolved experiments.

Following previous theoretical and experimental studies performed on trions \cite{esser_photoluminescence_2000,zipfel_electron_2022}, we assume here an exchange of energy between biexcitons and excitons \cite{bauer_optical_2013}. \blue{A simple derivation of the model is presented in the SM.} The excitonic energy levels involved in the radiative recombination of classical biexciton relaxations are represented in Fig. \ref{fig:1} (d). $E_{0}$ and $E_{XD}$ are the energy of the bright and dark exciton respectively.
In this classical biexciton recombination process (i.e. biexcitons with no kinetic energy) we would detect the PL emission energy $E_\gamma$ at the bright exciton energy reduced by the binding energy of the biexciton. In our case where a hot biexciton gas is assumed (i.e. high kinetic energy of the biexciton gas)  $E_\gamma$ is further reduced by the kinetic energy left to the dark exciton. The spectral asymmetry can thus be written as a convolution between a classical excitonic line represented by a Gaussian $g(E)$, a Heaviside function $\Theta(E_\gamma-E)$\blue{, and an exponential function }which reflects the low-energy broadening by the temperature $T$ \cite{park_imaging_2021,zipfel_electron_2022}:

\begin{equation}
\begin{aligned}
I(E) &\propto g(E) \otimes 
e^{\frac{E-E_\gamma}{\epsilon}}
\,\Theta(E_\gamma-E)
\\
\frac{1}{\epsilon} &=
\frac{M_{X_D}}{M_{X_0}}
\frac{1}{k_B T}
+ \alpha
\end{aligned}
\label{eq:recoil}
\end{equation}

where $M_{X_0}$ and $M_{X_D}$ are the bright and dark exciton masses, $E_\gamma$ is the PL peak position of the biexciton, and $k_B$ is Boltzmann's constant. The ratio $M_{X_D}/M_{X_0}$ reflects the mass ratio between the particle promoted at higher kinetic energy and the particle that disappears during the process ; we consider this ratio being equal to 1.18 \cite{kormanyos_k_2015,kapuscinski_rydberg_2021}. The coefficient $\alpha$, which does not depend on the temperature (i.e. constant over the excitation power), is linked to the oscillator strength \cite{esser_photoluminescence_2000,zipfel_electron_2022}. The temperature of the biexciton gas $T$ \blue{is} extracted from the above expressions and we will discuss \blue{its} spatial and temporal variations in the following.

\subsection{Spatial and spatio-temporal analysis}
\begin{figure*}
   \centering
   \includegraphics[width=0.9\linewidth]{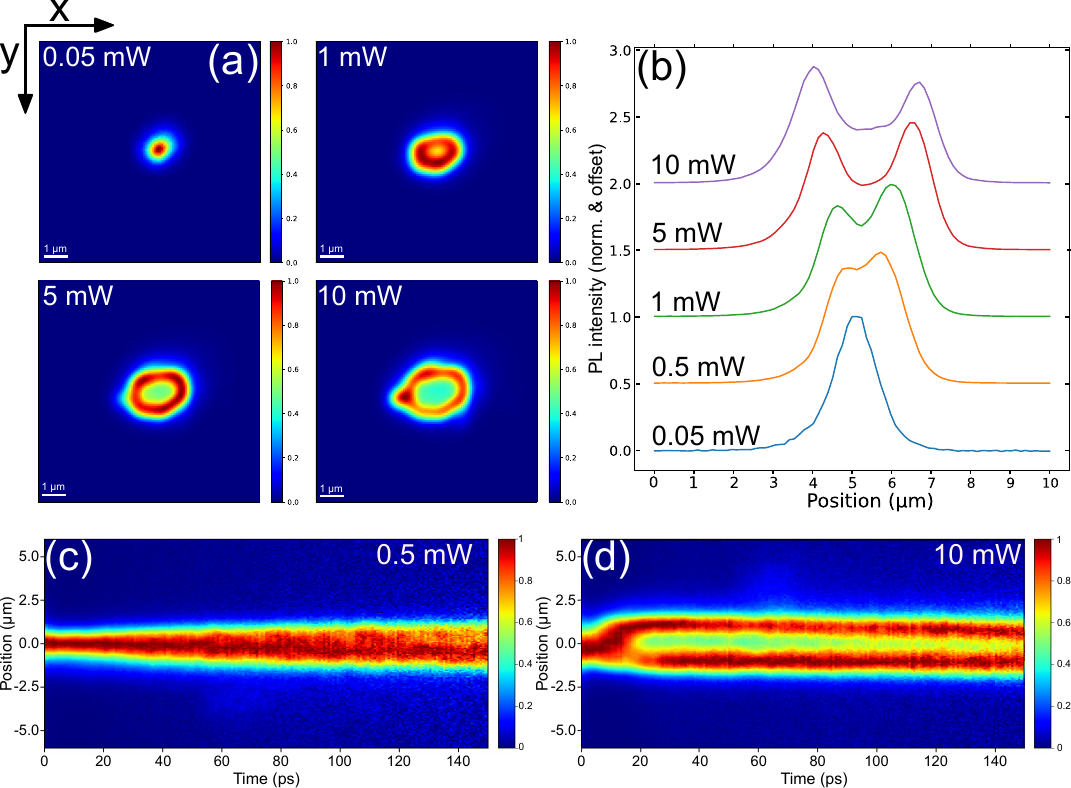}
    \caption{(a) Two-dimensional images of PL after excitation with the pulsed Ti-Sa laser at $700~nm$. Four average excitation power ranges are shown, from $50~\mu W$ to $10~mW$. (b) PL intensity profile normalized to the intensity maximum from a cross-section along the Y axis. Profiles are offset in intensity for clarity. (c) and (d) Time-resolved PL intensity profiles at $0.5~mW$ and $10~mW$. The spatial profiles are normalized for each time step. }
    \label{fig:2}
\end{figure*}

Figure \ref{fig:2}(a) shows biexciton \blue{photoluminescence} imaging results by spectrally selecting the PL emission from the biexciton with tunable edge filters. Excitation is still generated with the pulsed Ti:Sa laser at $700~nm$ and a CMOS camera is used for \blue{detection} with a spatial resolution of about $700~nm$. \blue{We use a lens with a focal length of 25$~cm$ in front of the camera giving rise to a magnification of  8.3 for our imaging system.} A general spatial broadening can be observed at low densities (see \blue{also SM Fig S2 for cw excitation}), but from $1~mW$ onwards, a progressive halo is observed which increases in size with increasing excitation power. This halo is characterized by the appearance of a double-peak structure in the PL spatial profiles (Figure \ref{fig:2}(b)). Biexcitons appear to be moving rapidly along the diameter of the halo, away from the central region. \blue{This halo is observable up to 60K (see SM, Fig S7).} Note that such halos can be observed in all areas of the sample \blue{and for the bright exciton PL emission (see SM, Fig. S3).}

 
This particular shape has already been observed in the literature for excitons \cite{perea-causin_exciton_2019,zipfel_exciton_2020,uddin_enhanced_2022} or trions \cite{sun_observation_2014}. For example, at high excitation density in a $WS_2$ monolayer, a halo-shaped spatial pattern was observed when imaging the transport of neutral excitons\cite{perea-causin_exciton_2019,zipfel_exciton_2020}. This phenomenon was also observed in $MoSe_2$ by looking at the trion gas transport\cite{park_imaging_2021}. In both cases, the microscopic mechanism governing the halo formation and the observed unconventional diffusion have been attributed to the formation of strong spatial gradients in the excitonic or trionic temperature. However, the halo has never been observed previously for heavier excitonic species such as biexcitons.

To better analyze the halo formation, we then observe the temporal evolution of the biexciton PL spatial profiles along the \blue{diameter} of the halo. Figure \ref{fig:2}(c-d) shows streak camera images representing time-resolved PL spatial profile for two excitation powers (at each time $t$ the intensity profile has been normalized to the maximum intensity), one below the minimal power condition for the halo formation at 0.5$~mW$ and one above at 10$~mW$. In the lowest power condition, we already observe a non-linear diffusion where the spatial expansion with time is seen in the first 50$~ps$ followed by a slower expansion. This trend is \blue{typically a signature of Auger effect which depletes the excitation center at early times, producing a rapid apparent spatial broadening that saturates once the density decreases} \cite{lamsaadi_kapitza-resistance-like_2023,zipfel_electron_2022,kulig_exciton_2018}. At the highest power condition, we observe a rapid expansion in the first tens of picoseconds, then a spatial separation into two branches (i.e. halo ring) followed by spatial stabilization over time. We discuss the mechanisms of halo formation in the following sections.

\subsection{Time-resolved PL spectra}
Using the streak camera coupled to a monochromator, we then measure biexciton PL spectra as a function of time (see SM, Fig S4), the luminescence is here spatially integrated. Based \blue{a global spectral fitting taking into account the the recoil effect for the biexciton line}, we were able to fit the spectra at different times $t$ in order to extract the temporal evolution of the biexciton temperature and PL peaks position shift (See SM, part VIII).

\begin{figure}
   \centering
   \includegraphics[width=1\linewidth]{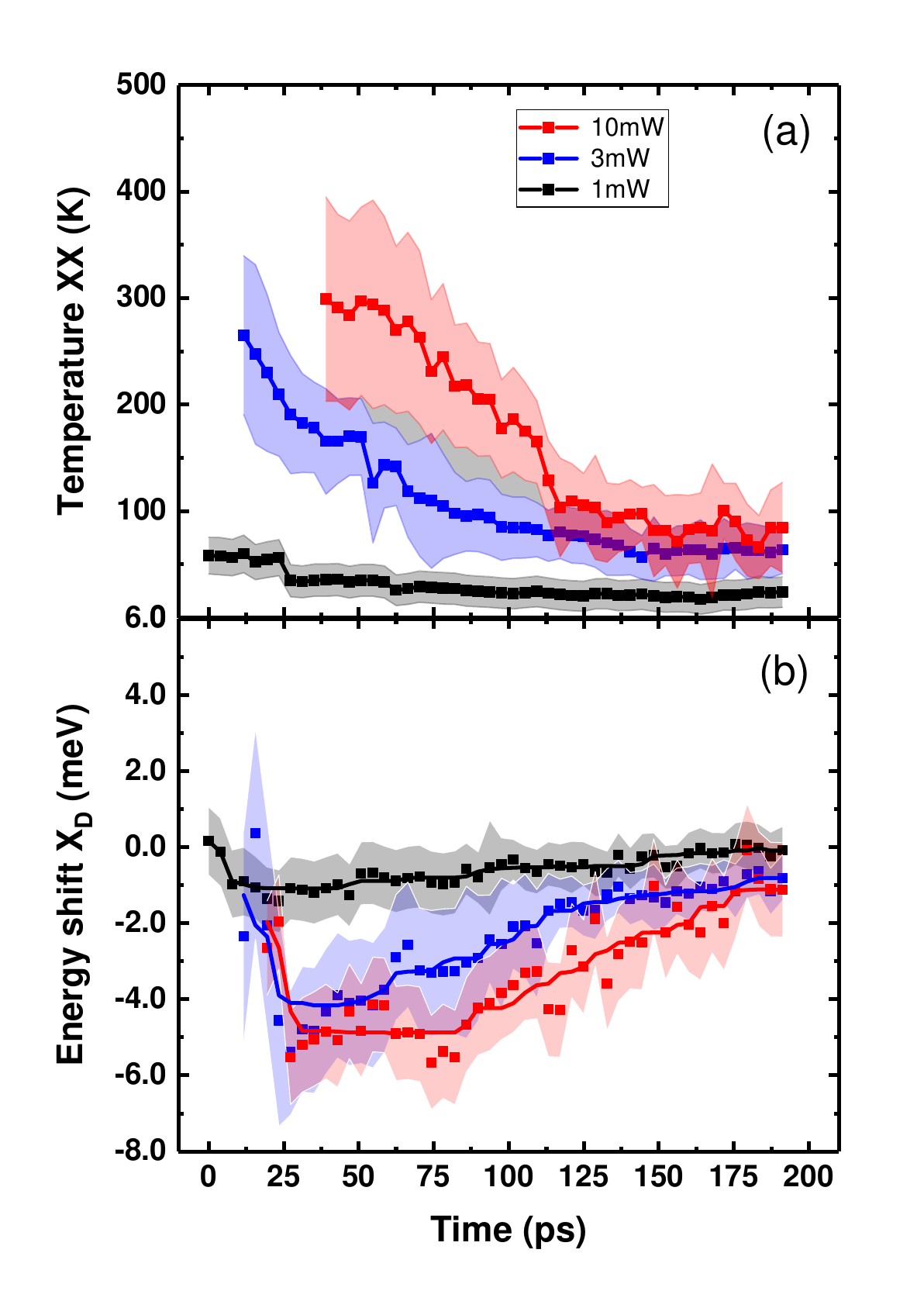}
    \caption{(a) time evolution of the biexciton temperature and (b) time evolution of the \blue{dark exciton} energy shift extracted from time resolved PL spectra for three different excitation powers\blue{; the lines are smoothed curves and serve as a guide to the eye.}}
    \label{fig:3}
\end{figure}

Figure \ref{fig:3} (a) shows the temporal evolution of the biexciton temperature for different excitation powers with the uncertainty indicated by the shaded colored areas \cite{uncertainty}.
We  observe a clear heating with increasing excitation power and an \blue{effective} decay time of temperature, \blue{in the order of $30-50~ps$ at the highest powers}.
\blue{At short times and high excitation power, the carrier density may exceed the Mott transition threshold, which questions the validity of the recoil model and the reliability of the extracted temperatures. We therefore avoid quoting temperature values during the first few picoseconds; further details are provided in Supplementary Material, Section VIII.}.
Figure \ref{fig:3} (b) shows the energy position shift of the \blue{dark exciton} as a function of time for the different excitation powers. \blue{This excitonic peak position is always symmetrical -as compare to the biexctonic line- which allows an accurate peak position extraction  reflecting the lattice heating.} At the lowest power, the position of the peak remains almost unchanged over time within the uncertainty of our method and of the experimental conditions. When the carrier density is sufficiently high (i.e. above 1$~mW$), we observe an energy shift of around $4~meV$ towards low energies in the first $20-30~ps$. Thereafter, the dynamics are reversed, with a progressive shift back to the initial energy with a time constant of about $50~ps$. 

In view of these experimental observations, we can suggest the following interpretation. As the density increases, Auger-type nonlinear effects  create, at the very first instants hot populations of quasi-particles (excitons, biexcitons at higher energies). These quasi-particles then thermalize, transferring their kinetic energy to the lattice, whose temperature rises during the first 25 picoseconds of the dynamics. The increase of lattice temperature is responsible for the low-energy shift of the \blue{dark exciton} peak position due to the decrease in the energy gap. The second phase of the kinetics (beyond $25~ps$) corresponds to lateral heat dissipation via the lattice, leading to a return to the initial energy of the \blue{dark exciton peak}, via a high-energy spectral shift (blue-shift). The lifetime of the biexciton, which is longer than that of the \blue{bright} exciton, makes it possible to track \textbf{biexciton gas temperature} over time. 

\blue{Concerning the spectral shift}, we can not rule out, at this stage, effects directly linked to exciton density. The decrease in repulsive exciton-exciton interactions as the exciton density decays with time is likely to lead to a redshift. This repulsive effect is mentioned by Park et al \cite{park_imaging_2021} to explain the spectral shifts in MoSe$_2$. In our case we could imagine similar effects whose decreasing efficiency with density would contribute to the redshift observed during the first \blue{tenth of} picoseconds. On the other hand, density-related effects also include a reduction of the energy gap with increasing density, due to the band gap renormalization (not perfectly compensated by the decrease of the binding energy). As the density decreases as a function of time, it will lead to a blueshift that could also contribute to the global blue spectral shift observed during the second phase of the kinetics.

\begin{figure}
   \centering
   \includegraphics[width=1\linewidth]{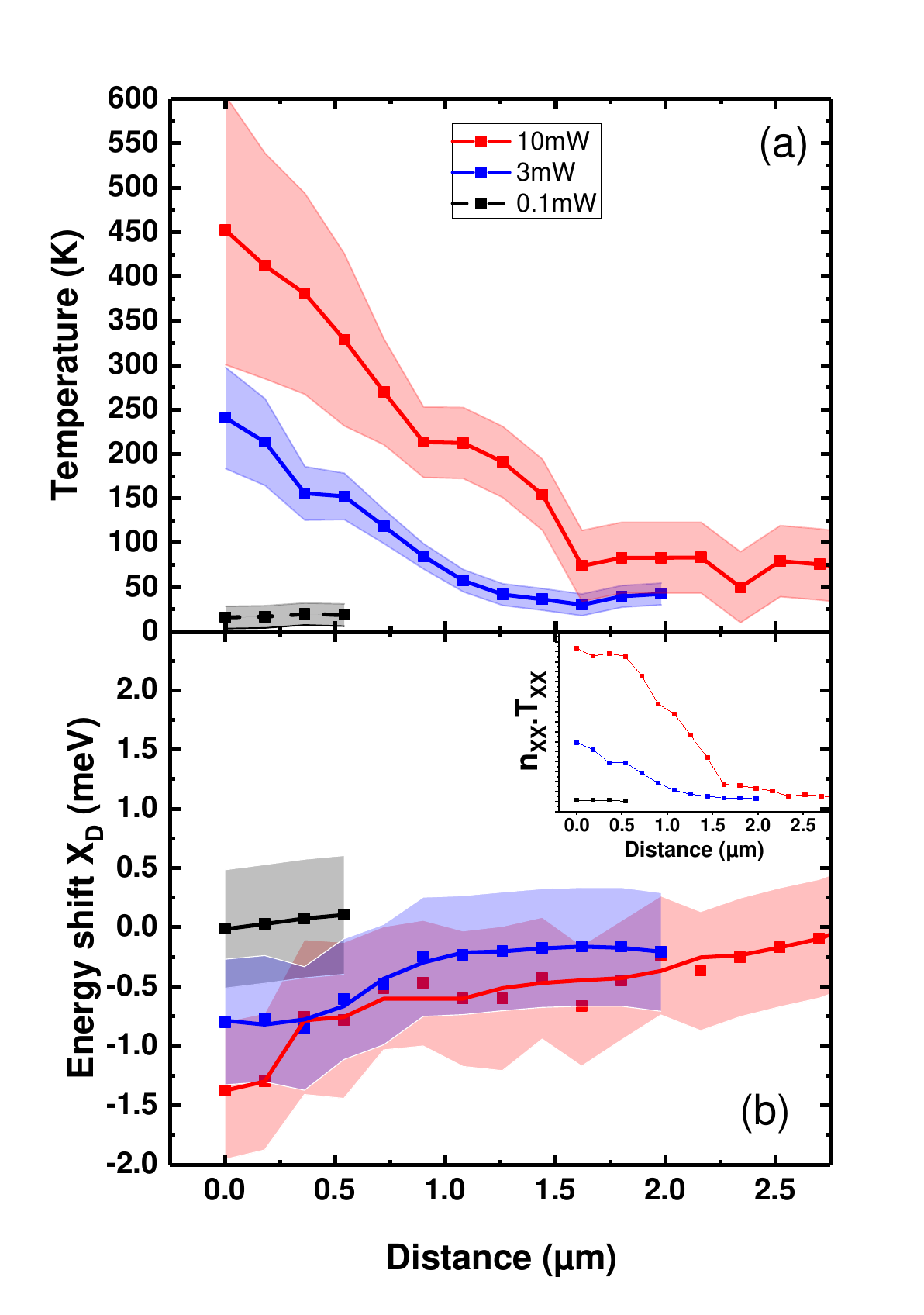}
    \caption{(a) spatial profiles of the biexciton temperature and (b) spatial profile of the \blue{dark exciton} peak energy shift extracted from energy-resolved PL profile (hyperspectral image) performed at 3 different excitation powers\blue{; the lines are smoothed curves and serve as a guide to the eye.} \blue{The inset shows the spatial evolution of the product of the biexciton density and temperature.}}
    \label{fig:4}
\end{figure}

To further confirm the presence of a hot biexcitonic gas, we then performed hyperspectral imaging (energy-resolved PL images along the diameter of the halo) where the halo is clearly visible at high excitation power (see SM, fig. S5). This allows us to analyze the spatial variation of the biexciton temperature and \blue{the dark exciton PL peak energy shift}. Note that this measurement has been carried out on a second similar sample. Following the previous fitting procedure, we extracted the spatial profiles for the biexciton temperature and the \blue{dark exciton} PL peak position, shown in Figures \ref{fig:4}(a) and (b) respectively. \blue{On the one hand}, the temperature profile \blue{clearly} reveals a higher temperature at the center of the halo. 
\blue{On the other hand, a red-shift is observed indicating that thermalization occurs gradually, warming up the lattice.}

\blue{The red-shift observed away from the excitation spot can be attributed to the interplay between heating induced by exciton thermalization and transport. The red-shift is expected to roughly scale with the product $n_{XX} . T_{XX}$, which is consistent with the behavior shown in the inset of Fig. \ref{fig:4}(b): a monotonic variation, with a slight flattening near the center.} \blue{Note that s}imilar trends are also observed for the trion in $MoSe_2$ in Ref \cite{park_imaging_2021}. 


\blue{Finally, to emphasize the key role of the hot biexciton gas in the transport, a classical model of the phenomenon is presented in the Supplementary Information. This model considers a biexciton temperature profile vanishing with a characteristic time $\tau_D$. The lateral current depends on the concentration gradient (i.e. pure diffusion) as well as the temperature gradient via the Seebeck coefficient $S$ : }

\begin{equation}
   \blue{\vec{j}=D\vec{\nabla \mu} + \frac{\sigma S}{q} \vec{\nabla T}}
   \label{current}
\end{equation}

\blue{With $\mu$ the chemical potential,  $\sigma=nq\mu_{ex}$ the conductivity which depends on the mobility $\mu_{ex}$, the elementary charge $q$ and $D=kT\mu_{ex}$ the diffusion constant in $eV/cm^2$.
The modeling is performed by solving the classical diffusion (continuity) equation \ref{eqdiff}, using the expression for the current given in Eq. \ref{current}.}

\begin{equation}
 \blue{\frac{\partial n}{\partial t} = G_n(t=0) - R_n - div~\vec{j} }
\end{equation}\label{eqdiff}

\blue{ We employ standard parameter values, $S=300~\mu V/K$, $\mu_{ex}=800~cm^2/(V.s)$, $\tau_D=30ps$ and $R_n=-n/\tau$  with $\tau=100ps$\cite{park_imaging_2021,zipfel_electron_2022}. With these parameters, an initial temperature of $T_0=500~K$ provides good agreement with the experimental data (see Supplementary Material). We further note that the halo formation occurs only for $T_{0}>150K$ and it vanishes below this threshold. This supports the interpretation that a hot biexciton gas transport is driven by the coexistence of a heat and diffusion currents, thereby leading to the emergence of the halo. }


\subsection{Conclusion}
We have studied excitonic transport in $WSe_2$ at high carrier densities and focused on the transport of biexcitonic species. Spatial measurements showed non-linear diffusion with increasing excitation power. A study of the biexciton photoluminescence signal as a function of space and time revealed the formation of a halo at sufficiently high exciton densities. This particular diffusion profile is linked to the creation of hot biexcitonic population, inducing a strong spatial temperature gradient. The energy transfer from hot biexciton population towards the lattice could explain the observed time-dependent spectral shift \blue{of the dark exciton line}. The relatively long biexciton lifetime compared to the exciton one allows tracking the dynamics of the observed phenomena. Finally, the measurement of the spatial variation of the temperature gas and PL shift are in agreement with the existence of a radial Seebeck current and thermalization effect. These findings highlight and confirm the role of high-energy populations in excitonic transport within TMD, even for complex and heavy excitonic particles. \\

%

%
%

\textbf{Acknowledgements}\\ This study has been partially supported through the EUR grant NanoX no. ANR-17-EURE-0009 in the framework of the \textit{Programme des Investissements d’Avenir}. We also thank the support of the French Agence Nationale de la Recherche with the funding under the project SOTspinLED (grant number ANR-22-CE24-0006-01), and the program ESR/EquipEx+ (grant number ANR-21-ESRE-0025).

\providecommand{\latin}[1]{#1}
\makeatletter
\providecommand{\doi}
  {\begingroup\let\do\@makeother\dospecials
  \catcode`\{=1 \catcode`\}=2 \doi@aux}
\providecommand{\doi@aux}[1]{\endgroup\texttt{#1}}
\makeatother
\providecommand*\mcitethebibliography{\thebibliography}
\csname @ifundefined\endcsname{endmcitethebibliography}
  {\let\endmcitethebibliography\endthebibliography}{}

\clearpage


\onecolumngrid

\setcounter{page}{1}
\setcounter{figure}{0}
\setcounter{table}{0}
\renewcommand{\thefigure}{S\arabic{figure}}
\renewcommand{\thetable}{S\arabic{table}}

\vspace*{2cm}
\begin{center}
    {\LARGE \textbf{Supplementary Materials : Thermal assisted transport of biexcitons in monolayer WSe$_2$}}\\[0.5cm]
    {Dorian Béret$^{1}$, Louka Hemmen$^{1}$, Vishwas Jindal$^{1}$, Sreyan Raha$^{1}$, Thierry Amand$^1$, Delphine Lagarde$^1$, Andrea Balocchi$^1$, Cédric Robert $^1$, Hélène Carrere$^1$, Xavier Marie$^1$, Pierre Renucci$^1$, Laurent Lombez$^1$}\\
    \textit{$^1$Universit\'e de Toulouse, INSA-CNRS-UPS, LPCNO, 135 Avenue Rangueil, 31077 Toulouse, France}\\[1cm]
\end{center}
\vspace{1cm}

\section{Complement to PL spectra power dependence}
\begin{figure}[h]
   \includegraphics[width=0.5\linewidth]{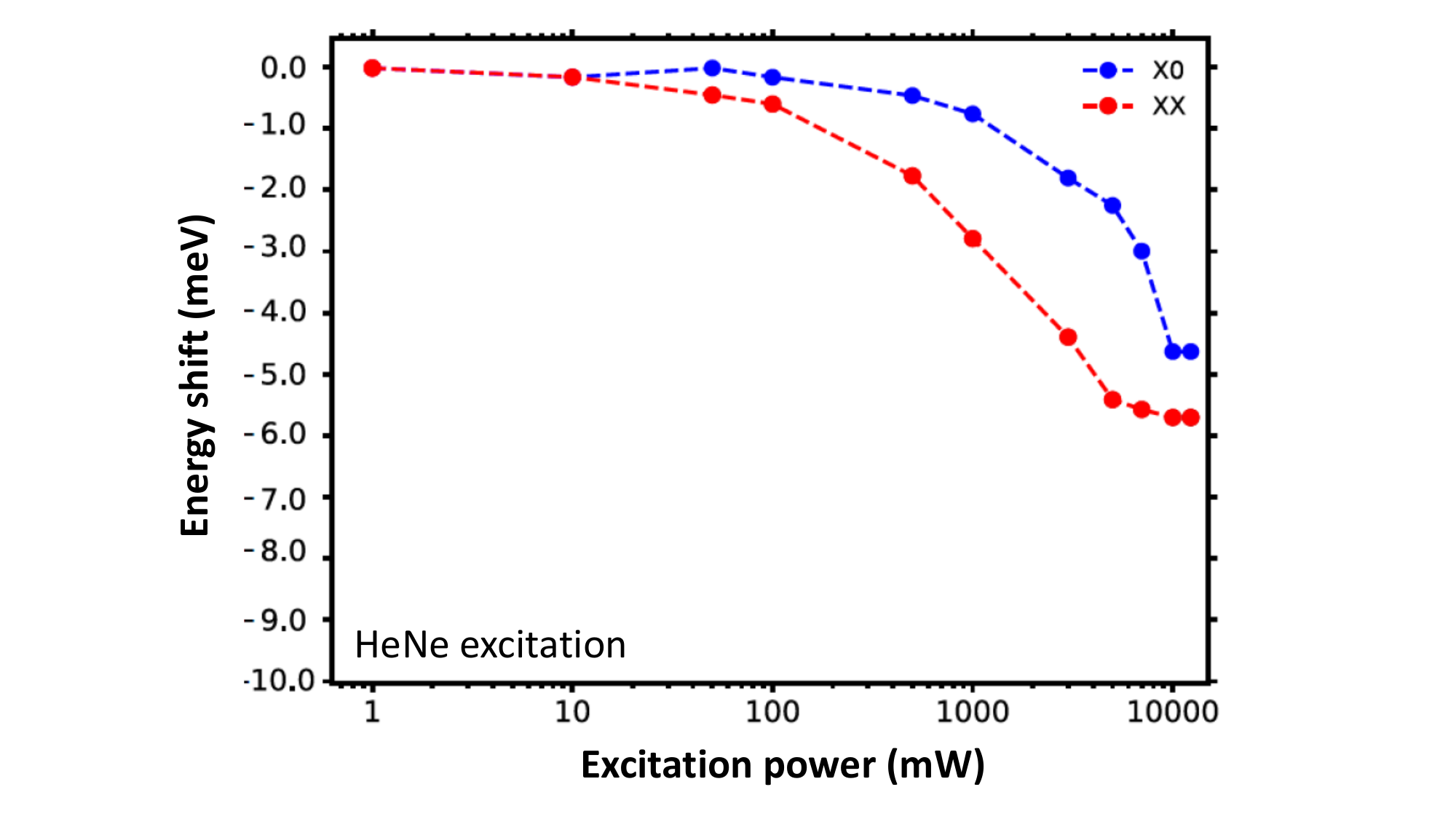}
  \caption{Evolution of the PL peak energy for the exciton (blue) and the biexciton (red)  as a function of the HeNe laser power. }
    \label{fig:SIpower}
\end{figure}

Figure \ref{fig:SIpower} presents the evolution of the PL peak energy for the exciton (blue) and the biexciton (red) as a function of the HeNe laser power.

\section{Biexciton PL spatial profile under cw H\lowercase{e}N\lowercase{e} excitation}
\begin{figure}[h]
   \includegraphics[width=1\linewidth]{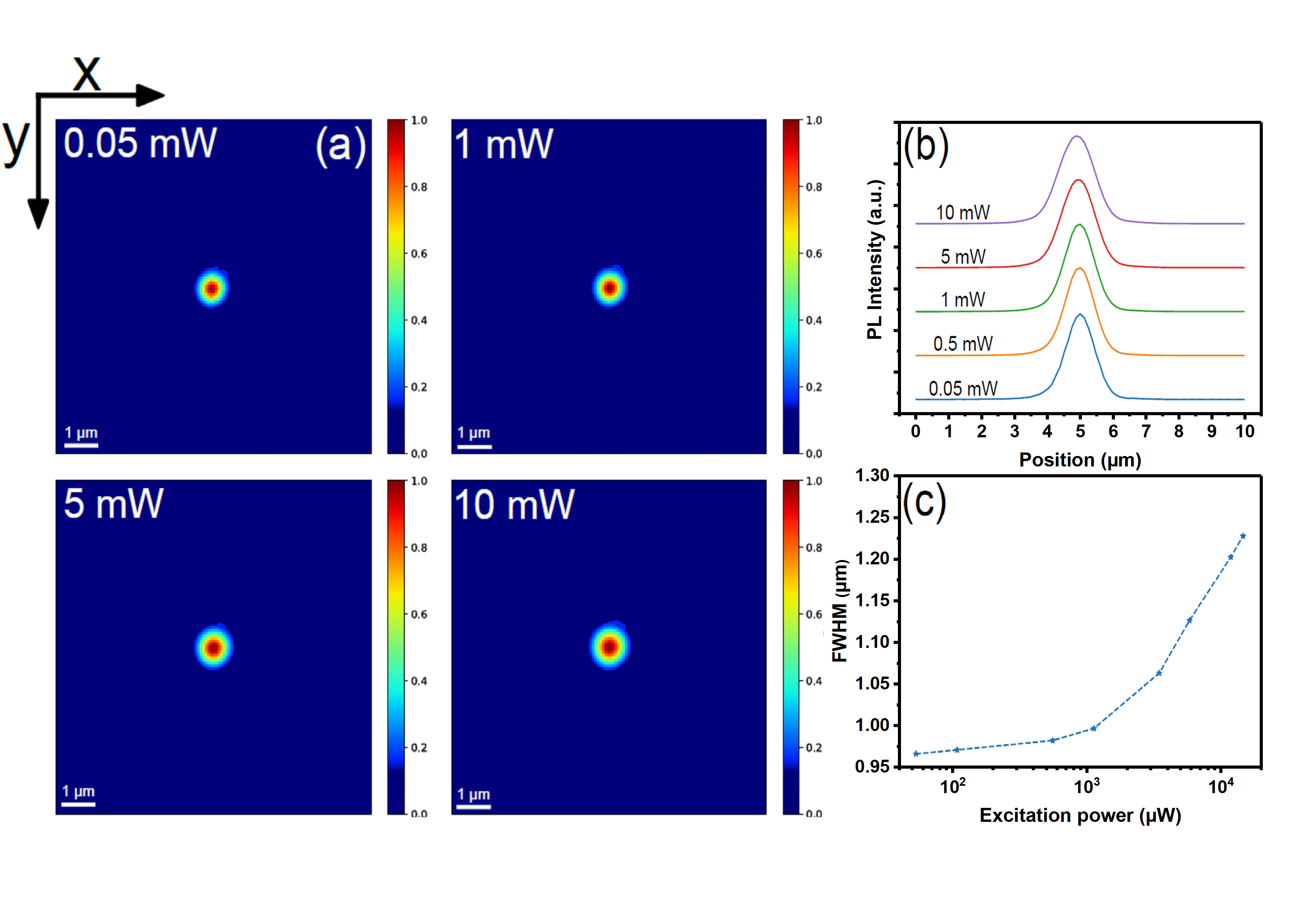}
  \caption{(a) Images of PL under focused cw HeNe excitation laser at four different excitation powers (b) extraction of four different PL profiles from (a) along the diameter of the halo in the y axis (c) \blue{spatial} broadening of the PL profile as a function of the excitation power. }
    \label{fig:SIHalo}
\end{figure}

Figure \ref{fig:SIHalo} shows PL images and spatial profiles under cw HeNe excitation laser with different powers. No halo is visible but a broadening of the PL spatial profile is observed (see Figure \ref{fig:SIHalo}(c) extracted from a Gaussian fit). 

\newpage

\section{Exciton photoluminescence images}
\begin{figure}[h]
   \includegraphics[width=0.6\linewidth]{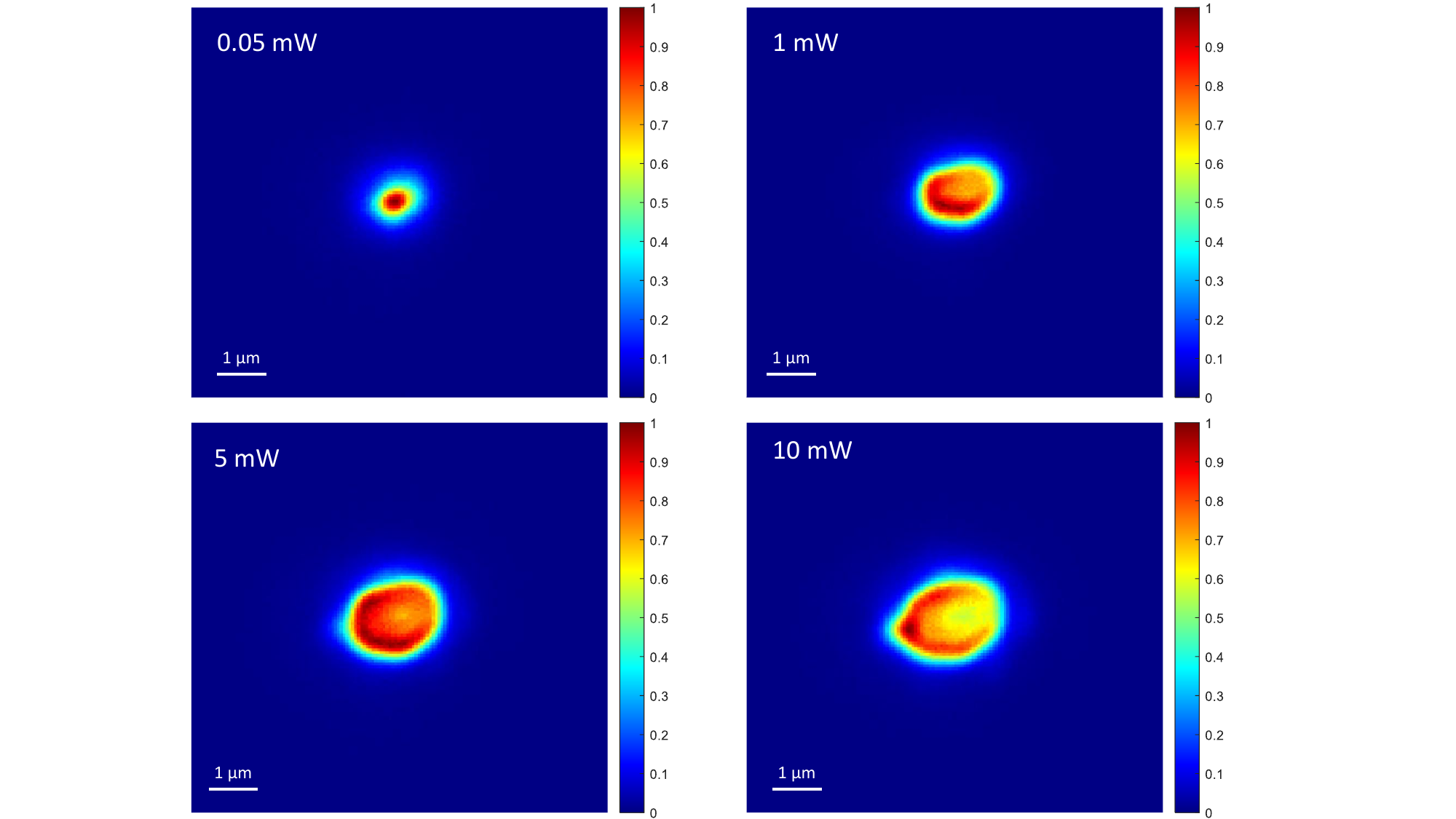}
  \caption{PL images of the exciton emission under $700~nm$ Ti:Sa \blue{pulsed} laser excitation with different powers.  }
    \label{fig:XHalo}
\end{figure}

Figure \ref{fig:XHalo} shows four PL images under pulsed $700~nm$ Ti:Sa laser excitation with different powers. The detection wavelength is centered around the exciton PL emission. As the exciton lifetime is short, it was not possible to temporarily resolved the corresponding halo, unlike in the case of the bi-exciton.

\newpage

\section{Time-resolved PL spectrum}
\begin{figure}[h]
   \includegraphics[width=1\linewidth]{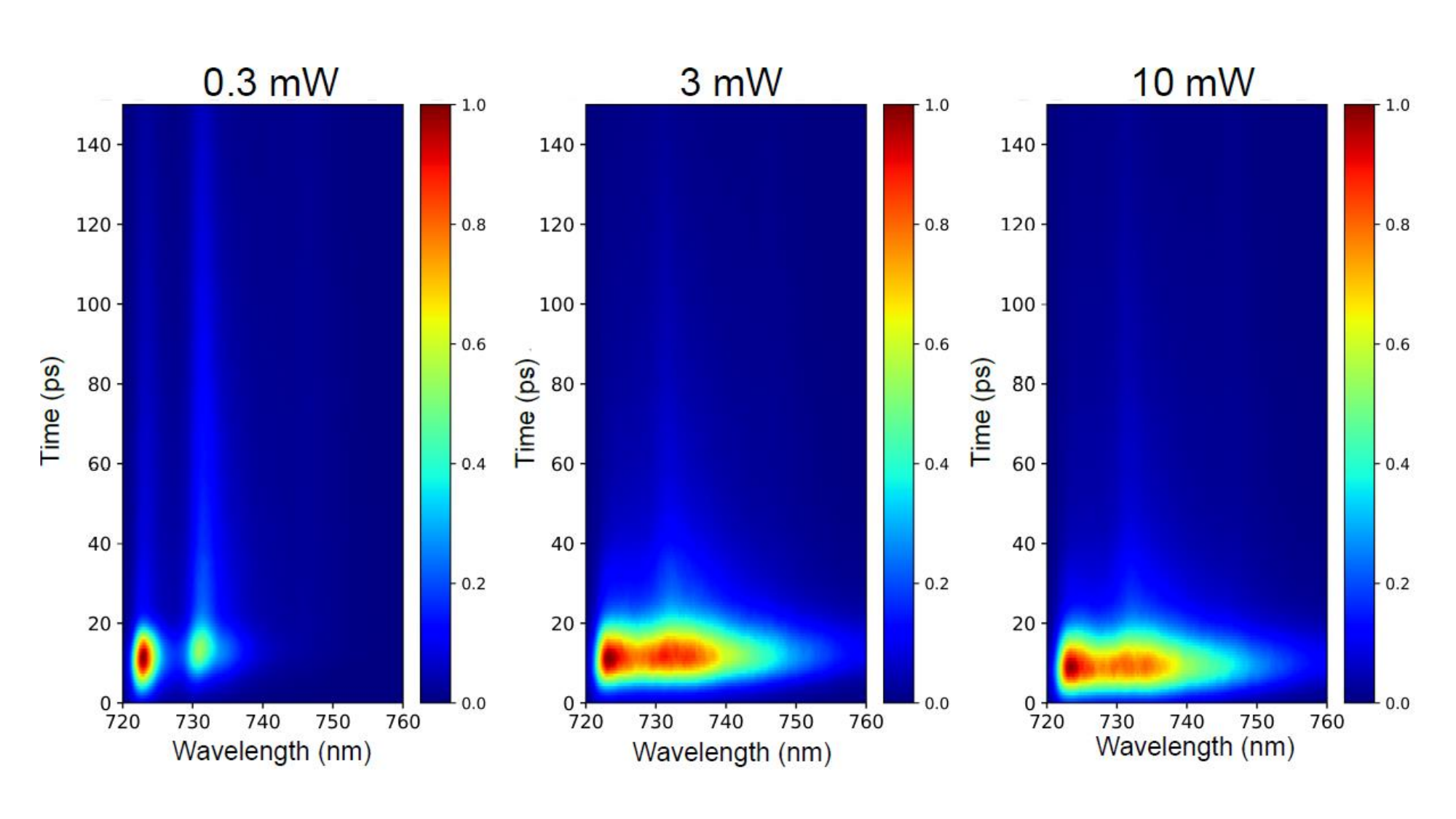}
  \caption{Time-resolved PL spectrum recorded for three laser excitation powers  }
    \label{fig:streak}
\end{figure}

Figure \ref{fig:streak} shows three time-resolved PL spectra recorded by the Streak camera at three excitation powers of the pulsed Ti:Sa laser set at $700~nm$ : $1~mW$, $3~mW$ and $10~mW$. 

\newpage

\section{Spectrally-resolved PL intensity spatial profile}
\begin{figure}[h]
   \includegraphics[width=1\linewidth,trim={0 5cm 0 5cm},clip]{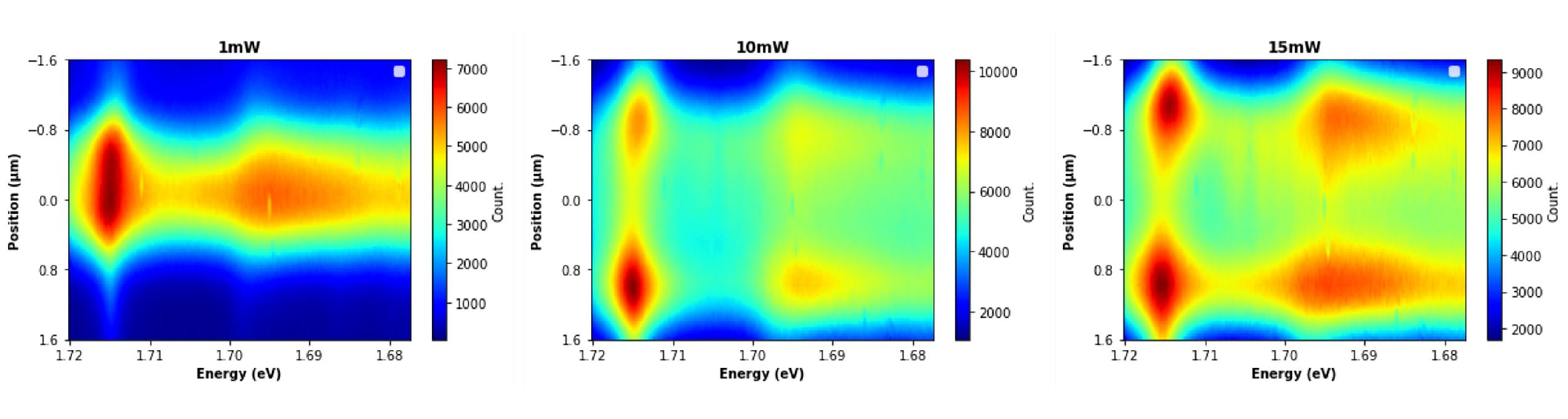}
  \caption{Hyperspectral images - Spectrally resolved PL intensity profile along a diameter of the halo for three excitation powers }
    \label{fig:imagesHI}
\end{figure}

Figure \ref{fig:imagesHI} shows three spectrally resolved PL intensity profiles along the diameter of the halo (hyperspectral images) recorded at three excitation powers of the pulsed Ti:Sa laser set at $700~nm$. Spatial profiles in Fig. 4 of the main text are values obtained \blue{from the positive side} of the halo.

\section{Reproducibility and temperature dependence}
\begin{figure}
   \includegraphics[width=1\linewidth]{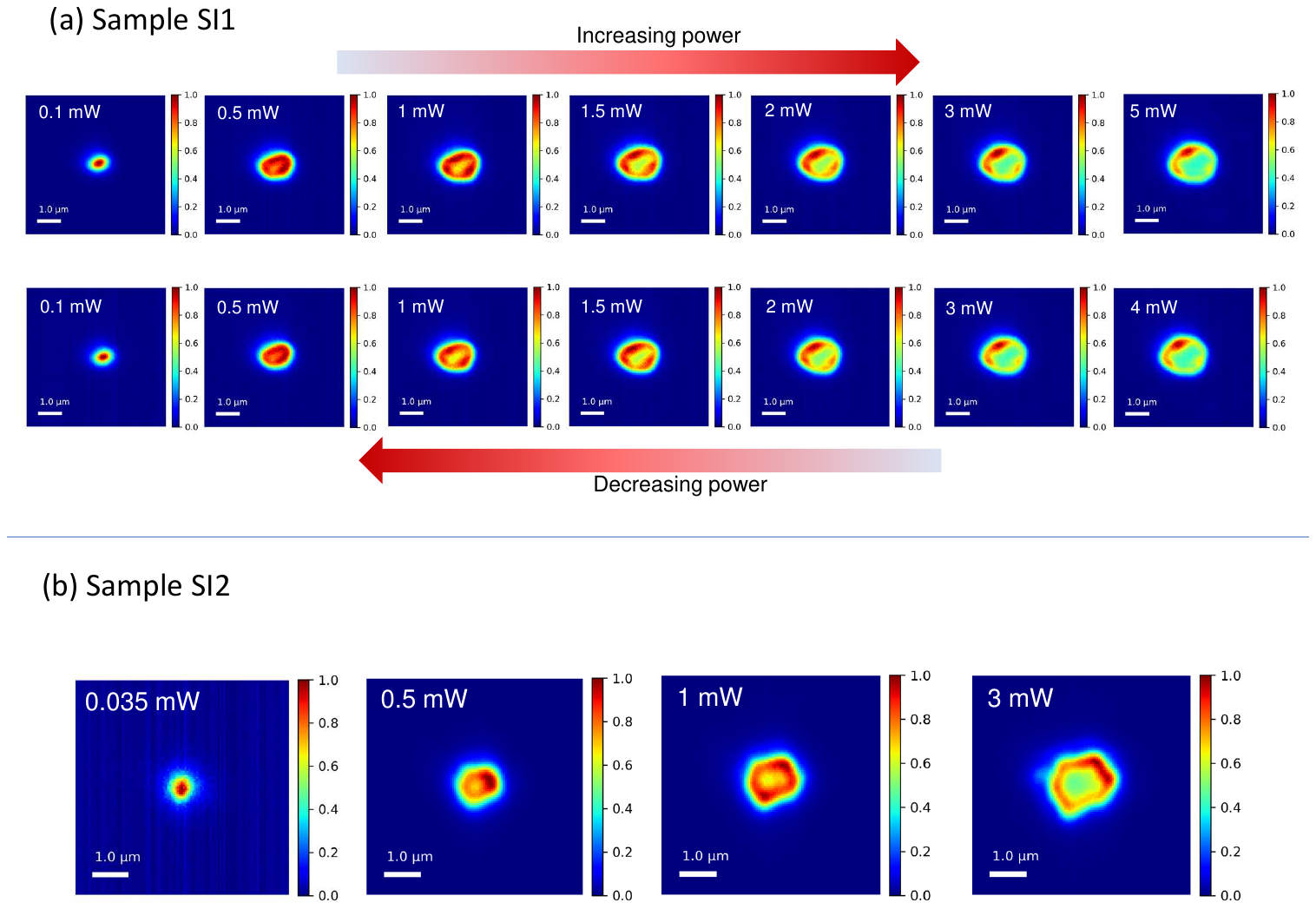}
  \caption{\blue{(a) Photoluminescence images acquired on sample SI1 while repeatedly switching between low and high power (b) Photoluminescence images acquired on sample SI2 to show the reproducibilty on another sample.}}
    \label{fig:repeat}
\end{figure}

\blue{In this section, we present results obtained on two additional samples, SI1 and SI2, different than the ones used in the main manuscript. Photoluminescence images of the biexciton emission from sample SI1 and sample SI2 are shown in Fig. \ref{fig:repeat}. PL images in Fig. \ref{fig:repeat} (a) illustrate the evolution of the halo while repeatedly switching between low and high power (700$nm$ Ti:Sa laser). PL images in Fig. \ref{fig:repeat}(b) are recorded at different power on sample SI2. The reproducibility is very good under repeated high-power excitation, with no observable sample degradation, and also across different samples.}\\

\begin{figure}
   \includegraphics[width=0.7\linewidth,trim={0 9cm 0 1cm},clip]{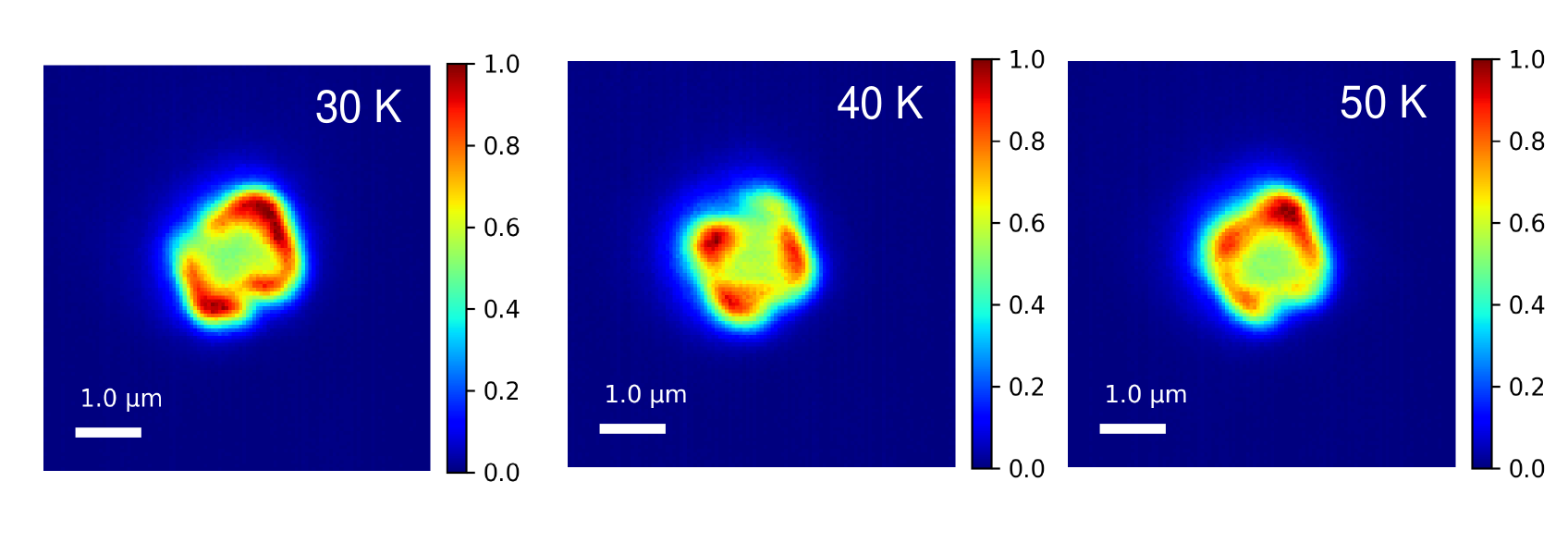}
  \caption{\blue{Photoluminescence images acquired on sample SI1 at various sample temperature. }}
    \label{fig:repeat2}
\end{figure}

\blue{Photoluminescence images acquired on sample SI1 are shown in Fig. \ref{fig:repeat2}, where the evolution of the halo is displayed as a function of the sample temperature. The halo remains observable up to 50 K from which the biexciton line can no longer be clearly identified. The corresponding PL spectra are also shown. }

\newpage

\section{Biexciton recoiling effect}
\blue{With a simple model we here explicit the expression of the PL Intensity $I(E)$ written in the main text : }

\begin{eqnarray}
    I(E) &\propto &  g(E) \otimes e^{\frac{E-E_\gamma}{\epsilon}}.\Theta(E_\gamma-E)
\\    1/\epsilon &=&  \frac{M_{X_D}}{M_{X_0}}\frac{1}{k_B.T} + \alpha    
\label{XX1}
\end{eqnarray}

\blue{where $M_{X_0}$ and $M_{X_D}$ are the bright and dark exciton masses, $E_\gamma$ is the PL peak position of the biexciton, and $k_B$ is Boltzmann's constant. The ratio $M_{X_D}/M_{X_0}$ reflects the mass ratio between the particle promoted at higher kinetic energy and the particle that disappears during the process ; we consider this ratio being equal to 1.18 \cite{kormanyos_k_2015,kapuscinski_rydberg_2021}. The coefficient $\alpha$, which does not depend on the temperature (i.e. constant over the excitation power), is linked to the oscillator strength \cite{esser_photoluminescence_2000,zipfel_electron_2022}. $T$ is the temperature of the biexciton gas and $E_\gamma$ is its the energy.}\\

\begin{figure}
   \includegraphics[width=0.4\linewidth]{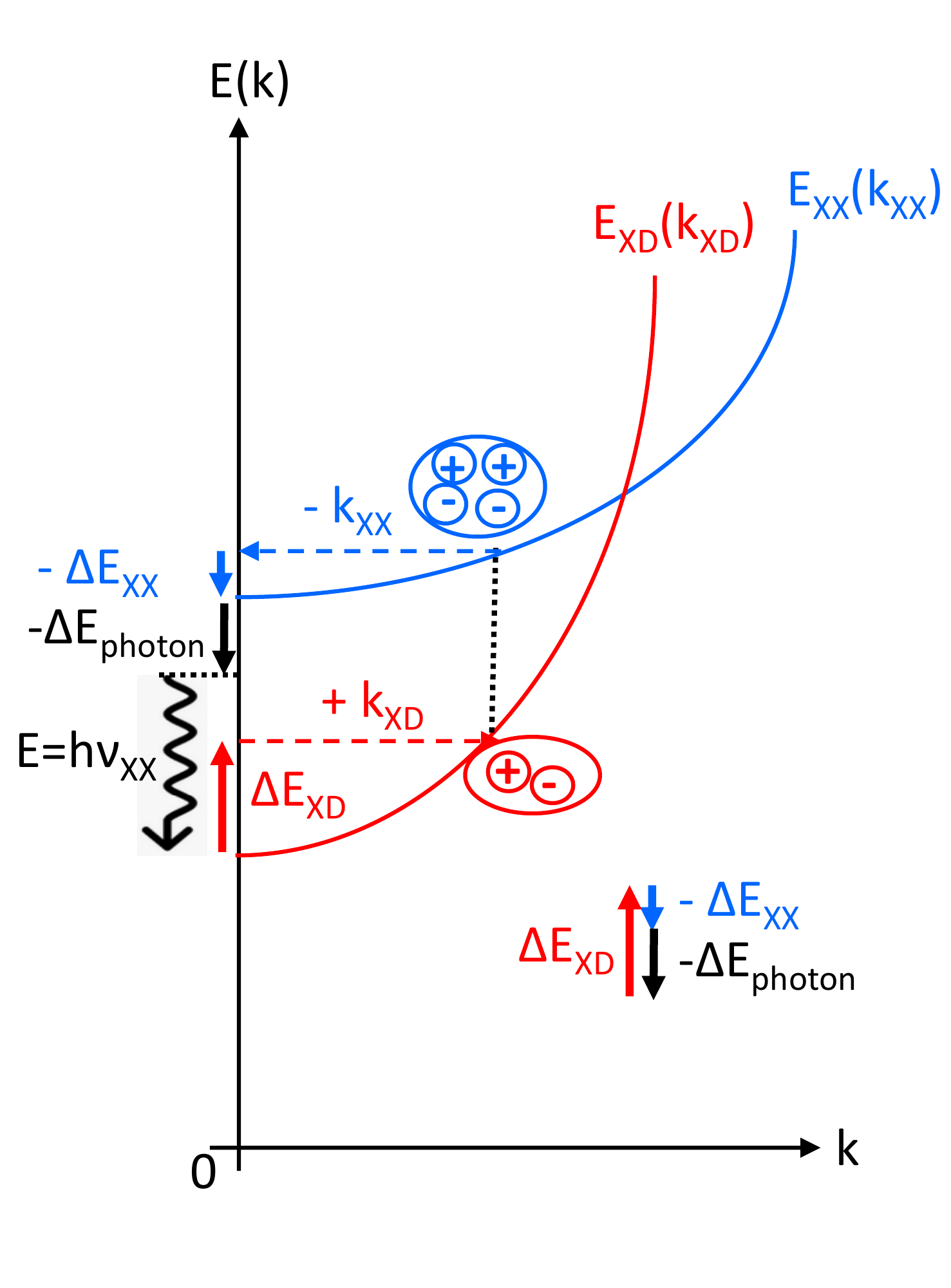}
  \caption{\blue{Diagram of the energies of the particles involved in the biexciton recoil effect. }}
    \label{petitmodel}
\end{figure}

\blue{The expression is based on the development that has been done on the trions \cite{zipfel_electron_2022}. The relevant energies are explicitly indicated in the schematic shown in Fig. \ref{petitmodel}. }

\blue{Let us begin by the energy of a biexciton  $E_{XX}(k)$ which is expressed by :
\begin{eqnarray}
   E_{XX}(k)&=&E_{X0} + E_{XD} - E_{binding}  + \frac{\hbar^2k^2}{2 M_{XX}}
\\ &=&   E_\gamma + E_{XD} + \frac{\hbar^2k^2}{2 M_{XX}}
   \label{XX2}
\end{eqnarray}
 with $E_{binding}$ is the binding energy of the biexciton, $k$ its wavevector, $M_{XX}$ its mass and $E_\gamma=E_{X0}- E_{binding} $. $E_\gamma$ is the photon energy corresponding to the biexciton PL peak.}

\blue{The biexciton recombination leads to photon with energy $E$ and dark exciton. Due to wavevector conservation, and neglecting the photon  wavevector, the dark exciton will acquire the wavevector of the biexciton. Its energy writes : 
\begin{eqnarray}
   E_{XD}(k)= E_{XD} + \frac{\hbar^2k^2}{2 M_{XD}}
   \label{XX3}
\end{eqnarray}}

\blue{Due to the energy conservation during the recombination process, one can write :
\begin{eqnarray}
   E_{XX}(k)= E + E_{XD}(k) 
   \label{XX3}
\end{eqnarray}}

\blue{So it leads to :
 \begin{eqnarray}
    E_\gamma + E_{XD} + \frac{\hbar^2k^2}{2 M_{XX}} = E+E_{XD} + \frac{\hbar^2k^2}{2 M_{XD}}
   \label{XX4}
\end{eqnarray}}
 
\blue{ which eventually gives : 
 \begin{eqnarray}
   -\frac{\hbar^2k^2}{2 M_{XX}}  = (E - E_\gamma)\frac{M_{XD}}{ M_{X0}}
   \label{XX5}
\end{eqnarray}}

\blue{The spectral evolution of the PL intensity can be written as \cite{esser_photoluminescence_2000, zipfel_electron_2022}:
 \begin{eqnarray}
  I(E) &\propto & g(E)\otimes \left( M^2 . exp(\frac{-\hbar^2k^2}{2 k_b T M_{XX}}).\Theta(E_\gamma-E)\right) 
   \label{XX6}
\end{eqnarray}}

\blue{where $M$ is the transition matrix element which reflects the oscillator strength of the transition and $k_b$ is the Boltzmann constant. For simplicity this matrix element is approximated by an exponential function $M^2 \propto exp(\alpha(E-E_\gamma))$ \cite{esser_photoluminescence_2000, zipfel_electron_2022, christopher_long_2017}.}

\blue{Therefore the energy dependence of the PL intensity writes :
 \begin{eqnarray}
 I(E) &\propto & g(E)\otimes\left( exp(\alpha).exp\left( \frac{E-E_\gamma}{k_b T}.\frac{M_{XD}}{M_{X0}}\right).\Theta(E_\gamma-E) \right) 
 \\ &\propto & g(E) \otimes e^{\frac{E-E_\gamma}{\epsilon}}.\Theta(E_\gamma-E)
   \label{XX7}
\end{eqnarray}}

\section{Fitted curves to extract the temperature and energy shift variations}

\blue{We here discuss the validity of the biexciton temperature estimation based on the recoil effect in view of two points: (i) the possibility that the biexciton density exceeds the Mott transition threshold, and (ii) the overall quality of the fits.}\\

\blue{The determination of the exact photogenerated density is always challenging. The main uncertainty lies in the absorption coefficient or the excitation energy/wavelength of the laser (~700$~mW$). To our knowledge there is no experimental data in the literature for hbN encapsulated WSe$_2$ monolayers. This excitation energy lies between the 1 s and 2s exciton energies where rather weak absorption is expected, in the range of 0.1-1\%. Below, we estimate the biexciton density in the worst case of 1\% absorption which we believe unlikely since very weak energy density od states is present between the 1s and 2s state.}\\
\blue{At an excitation power of 1$~mW$ (resp. 10$~mW$) -where the halo becomes visible- the pulse energy is 1.2$~pJ$ (resp. 12$~pJ$), corresponding to an estimated photogenerated exciton density (since the excitation is sub-bandgap and does not generate free electron–hole pairs) of approximately 5.2 10$^{13}~cm^{-2}$ (resp. 5.2 10$^{14}~cm^{-2}$). However, on a very short timescale of a few picoseconds, this density is expected to decrease significantly due to several processes: biexciton formation (introducing roughly a factor of 2 reduction), Auger recombination (with a coefficient on the order of 1$~cm^{-2}/s$, leading to about one order of magnitude decrease within a few picoseconds), and non-radiative recombination channels. This overall reduction in population -which in turn weakens screening and phase-space filling effects- is further supported by the fact that excitonic spectral lines remain distinguishable.}\\
\blue{A rigorous determination of the carrier density remains challenging and subject to significant uncertainties. Moreover, a direct correlation between the total PL intensity and the carrier density is very difficult due the existence  of several species with different radiative lifetime. This point is related to whether we are above the Mott transition and whether it is appropriate to extract parameters in that regime. For this reason, we choose not to report parameters extracted from the recoil model of the biexciton gas when the PL intensity exceeds 20\% of that measured at the maximum intensity (at t=0) at 10$~mW$ excitation power. In other words, we therefore report parameters extracted for delays beyond approximately 7$~ps$ (resp. 35$~ps$) for excitation powers of 1$~mW$ (resp. 10$~mW$).  } \\

\begin{figure*}
   \centering
   \includegraphics[width=1\linewidth]{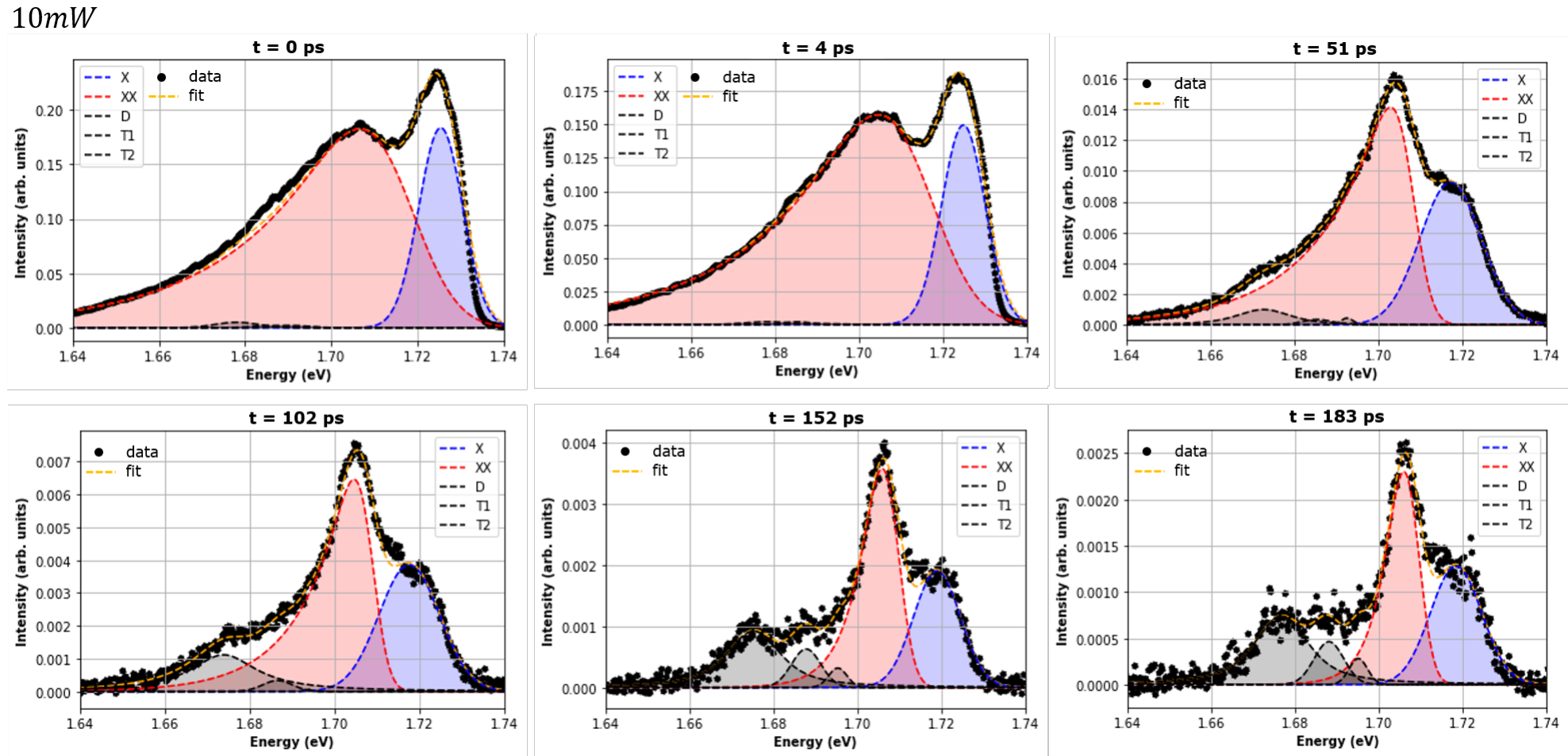}
  \caption{\blue{Fit and experimental data of time resolved spectra obtained at 10$~mW$ excitation power.}}
    \label{fig:Fit-time-10mW}
\end{figure*}

\begin{figure*}
   \centering
   \includegraphics[width=1\linewidth]{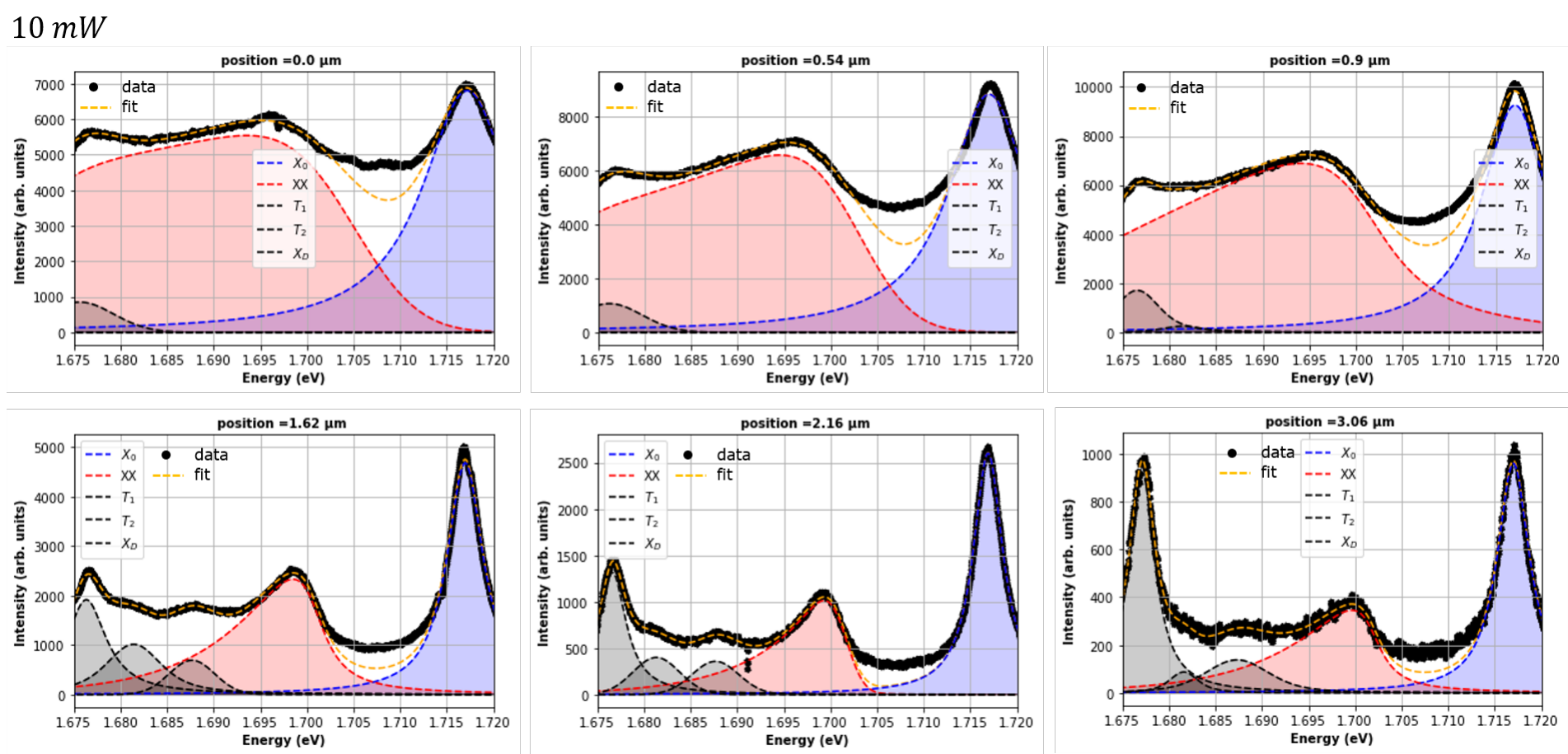}
  \caption{\blue{Fit and experimental data of spatially resolved spectra (hyperspectral measurement) obtained at 10$~mW$ excitation power.}}
    \label{fig:Fit-time-3mW}
\end{figure*}

\blue{Concerning the fitting quality, we have included examples of fits at the highest excitation power 10$~mW$ for time and spatially resolved spectra (see Fig. \ref{fig:Fit-time-10mW} and Fig. \ref{fig:Fit-time-3mW}). These examples show how the fitting procedure treats all the excitonic peaks. Interestingly, only the biexciton line is intrinsically affected by the asymmetry associated with the recoil-induced temperature increase. The comparison between data and fits shows overall good agreement across the full density range, both for time-resolved and spectrally-resolved data, supporting the robustness of the procedure. }

\section{Numerical modeling of halo}
As mentioned in the literature, halo formation can be induced by a spatial gradient in excitonic temperature. The latter will generate a current (i.e. particle flux) proportional to the temperature gradient $\nabla T$ (i.e. Seebeck current), which is added to the current induced by the particles gradient $\nabla n$, it writes \cite{callen_thermodynamics_1991}:

\begin{equation}
   \vec{j}=D\vec{\nabla \mu} + \frac{\sigma S}{q} \vec{\nabla T}
\end{equation}
With $\mu$ the chemical potential, $S$ the Seebeck coefficient, $\sigma=nq\mu_{ex}$ the conductivity which depends on $\mu_{ex}$ the mobility, $D=kT\mu_{ex}$ the diffusion constant in $eV/cm^2$.\\

It is possible to model a particle diffusion equation that takes the Seebeck effect into account, provided we know the spatial evolution of the temperature and its temporal evolution. \blue{It} has been modeled in the literature via a determination of $T(x,t)$ by the equipartition theorem, where the energy of the system is calculated by taking into account exciton-phonon interactions \cite{perea-causin_exciton_2019}.
More phenomenological modeling can be achieved by determining the coupled particle flux and heat flux. The heat flux is written as \cite{callen_thermodynamics_1991}:

\begin{equation}
   \vec{j_Q}=\sigma S \vec{\nabla \mu} + (T\sigma S^2 + \kappa)\vec{\nabla T}
\end{equation}
Where $\kappa$ is the thermal conductivity. 

 We then turn to the continuity equations for particles and heat.
\[
   \left\{
   \begin{aligned}
\frac{\partial n}{\partial t} &= G_n - R_n - div~\vec{j}\\ 
\frac{\partial (T.c_{el})}{\partial t} &= G_q - R_q - div~\vec{j_q}\\ 
   \end{aligned}
   \right.
\]

With $c_{el}$ the heat capacity, $G_n$ and $G_q$ being particle and heat generation, and $R_n$ and $R_q$ being particle recombination and heat loss.
The difficulty of this modeling lies in the concentration and temperature dependence of the various constants in the problem. Excitonic and phononic contributions in the heat current could also be separated \blue{in a more detailed model}.\\

\begin{figure*}
   \centering
   \includegraphics[width=1\linewidth]{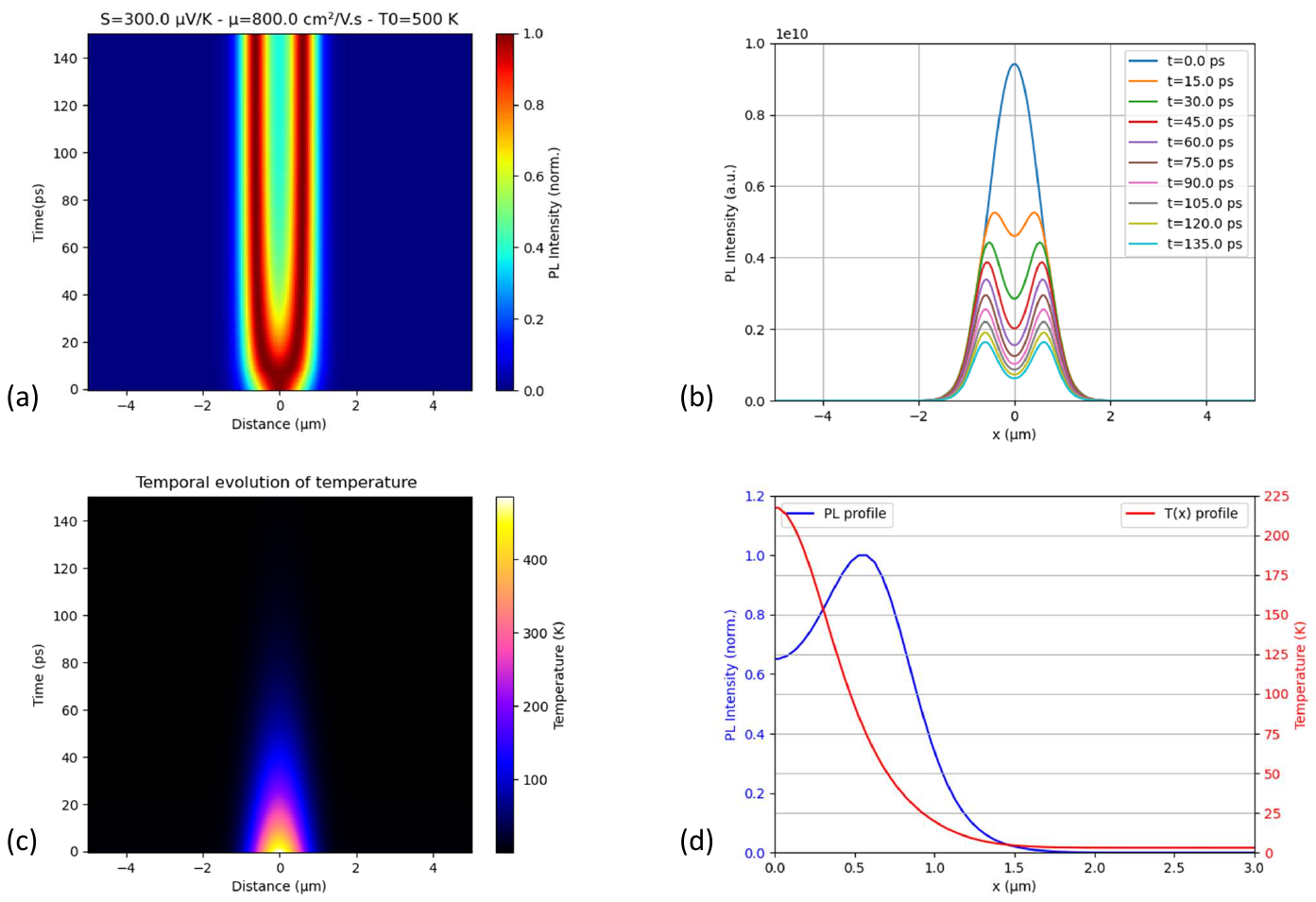}
  \caption{Modeling for $T_{0}=500K$ (a) Mapping of the temporal evolution of the intensity normalized by the intensity maximum at each time,(b) Photoluminescence intensity profiles at different times, (c) Mapping of the temporal evolution of the temperature profile, (d) Time integrated PL intensity profile and temperature. }
    \label{fig:SI1}
\end{figure*}

\begin{figure*}
   \centering
   \includegraphics[width=1\linewidth]{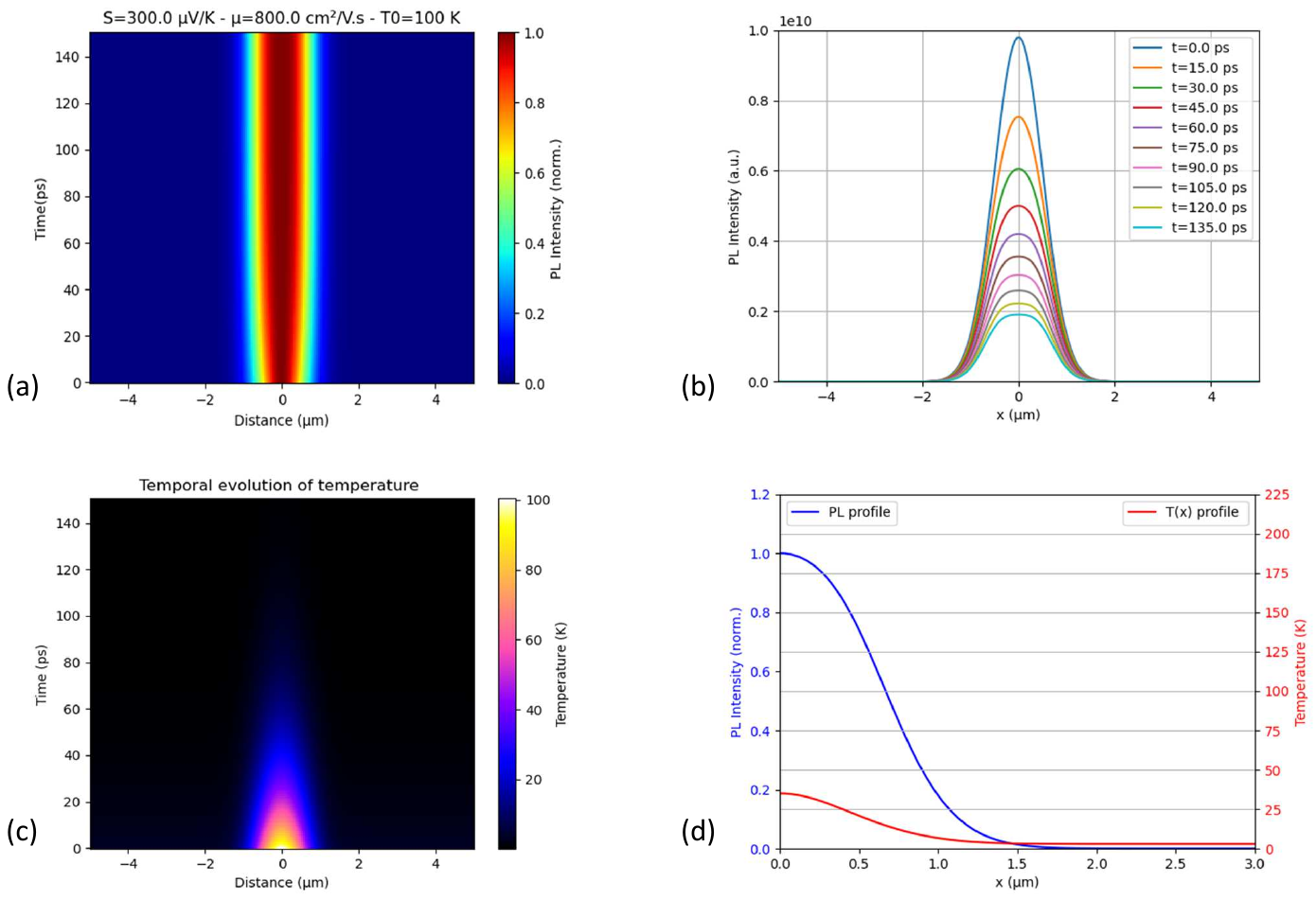}
  \caption{Modeling for $T_{0}=100K$ (a) Mapping of the temporal evolution of the intensity normalized by the \blue{spatial} intensity maximum at each time,(b) Photoluminescence intensity spatial profiles \blue{along the diameter} at different times, (c) Mapping of the temporal evolution of the temperature profile, (d) Time-integrated PL intensity and temperature spatial profiles. }
    \label{fig:SI2}
\end{figure*}

An alternative way that we will use in the following is to model in a simple manner the halo effect by assuming a spatio-temporal evolution of the biexciton temperature as :

\begin{eqnarray}\label{eq:Txt}
T(x,t) \propto G_n \propto T_0~exp(-x^2/2w_0^2)~exp(-t/\tau_Q)
\end{eqnarray}

with $T_0$ an initial maximum temperature, $w_0=0.5 \mu m^2$ defines the spatial broadening, and $\tau_Q=30~ps$ a characteristic time reflecting thermal dissipation as observed from time-resolved spectra from which we plot biexciton temperature as a function of time. \\
The results from the numerical calculations are shown in figure \ref{fig:SI1}. 
The modeling is carried out with classical parameter values \cite{park_imaging_2021,zipfel_electron_2022} such as $S=300~\mu V/K$ and $\mu_{ex}=800~cm^2/(V.s)$ and $R_n=-n/\tau$  with $\tau=100~ps$. With these parameters, the results are shown in Fig. \ref{fig:SI1}(a) for $T_0=500~K$ where we clearly see an agreement with the experimental results shown in Fig. 2 in the main text. Results obtained for low heating at $T_0=100~K$ (no halo) are also displayed in Fig. \ref{fig:SI2} for comparison, the halo phenomenon occurs when $T_{0}>150~K$.
For information, the PL intensity spatial profiles as a function of time are shown in \ref{fig:SI1}(b).
The time resolved temperature profile described by equation \ref{eq:Txt} is shown in figure \ref{fig:SI1}(c). 
Finally we also show the time integrated PL profile (blue curve in \ref{fig:SI1}(d)) and the time integrated temperature profile weighted by the PL intensity (red curve in \ref{fig:SI1}(d)). This simple modeling illustrates that the spatial halo phenomenon can be induced by a temperature gradient. It also underlines \blue {a bi-}excitonic population much warmer than the lattice.


\begin{mcitethebibliography}{34}
\bibitem[unc()]{uncertainty}
Note on uncertainties in {Fig} 3 and 4 :. The uncertainty $\Delta T\approx
  \frac{k_B.M_{X_B}}{M_{X_D}}T^2.\Delta \alpha$ is linked to the uncertainty on
  the estimation of the constant $\alpha$, it is nonlinear with $T$.\relax
\mciteBstWouldAddEndPunctfalse
\mciteSetBstMidEndSepPunct{\mcitedefaultmidpunct}
{}{\mcitedefaultseppunct}\relax
\EndOfBibitem
\end{mcitethebibliography}
\end{document}